\documentclass[prl,twocolumn,showpacs,preprintnumbers,amsmath,aps]{revtex4}
\usepackage{graphicx,hyperref}
\usepackage{amssymb}
\usepackage{bm}
\usepackage{subfigure}

\begin{document}

\title{Spinodals with Disorder: from Avalanches in Random Magnets to Glassy Dynamics}

\author{Saroj Kumar Nandi} \email{sarojnandi@gmail.com}
\affiliation{IPhT, CEA/DSM-CNRS/URA 2306, CEA Saclay, F-91191 Gif-sur-Yvette Cedex, France}

\author{Giulio Biroli} \email{giulio.biroli@cea.fr}
\affiliation{IPhT, CEA/DSM-CNRS/URA 2306, CEA Saclay, F-91191 Gif-sur-Yvette Cedex, France}

\author{Gilles Tarjus} \email{tarjus@lptmc.jussieu.fr}
\affiliation{LPTMC, CNRS-UMR 7600, Universit\'e Pierre et Marie Curie,
bo\^ite 121, 4 Pl. Jussieu, 75252 Paris c\'edex 05, France}

\date{\today}

\begin{abstract}
We revisit the phenomenon of spinodals in the presence of quenched disorder and develop a complete theory for it. 
We focus on the spinodal of an Ising model in a quenched random field (RFIM), which has applications in many areas from materials to social science. By working at zero temperature in the quasi-statically driven RFIM, thermal fluctuations are eliminated and one can give a rigorous content to the notion of spinodal. We show that the latter is due to the depinning and the subsequent expansion of rare droplets. We work out the associated critical behavior, which,  in any finite dimension, is very different from the mean-field one: the characteristic length diverges exponentially and the thermodynamic quantities display very mild nonanalyticities much like in a Griffith phenomenon. From the recently established connection between the spinodal of the RFIM and glassy dynamics, our results also allow us to conclusively assess the physical content and the status of the dynamical transition predicted by  the mean-field theory of glass-forming liquids.
\end{abstract}

\maketitle

The instability or failure of a metastable state in the presence of quenched disorder plays a key role in a wide spectrum of out-of-equilibrium situations, where it leads to extreme events such as macroscopic avalanches, shocks, ruptures or crises. Such ``spinodal transitions'' can be found for instance in the hysteretic behavior of disordered magnets \cite{bertotti,sethna93}, capillary condensation in disordered mesoporous materials \cite{aerogel14,detcheverry}, failure in amorphous materials \cite{failure,wyart14}, and in a variety of social and economic phenomena \cite{JP}. Whereas spinodal transitions are well understood in ``clean'' systems, where the limit of stability of a thermodynamically metastable state is associated with the emergence of ``soft modes'' \cite{binder,klein,debenedetti}, quenched disorder introduces new phenomena and drastically changes the physical behaviour: it pins the soft modes so that the physics is dominated by avalanches and bursts in place of long wave-length fluctuations.\\
In this work we present a complete theory of spinodal transitions in the presence of quenched disorder. In addition to its wide relevance from materials to social science, one main motivation to focus on this problem is its connection with glassy dynamics \cite{MCT-spinodal}, as unveiled by recent advances in the theory of glass formation.  Indeed, in what now appears to be the proper mean-field (MF) theory of the glass transition \cite{zamponi-HS,parisi-mezard},  the liquid as it is cooled first undergoes a dynamical transition, which is essentially that described by the mode-coupling theory (MCT) \cite{gotze,parisi-mezard}. This transition is found to be in the same universality class as the spinodal of the random-field Ising model (RFIM) \cite{MCT-spinodal}. The fact that the dynamical transition is the point where metastable states emerge (or loose their stability if one comes from low temperature) and that these states correspond to disordered particle arrangements provides the physical basis for this result. 
From field-theoretical arguments, the upper critical dimension above which the critical exponents take their MF value, \textit{i.e.}, here, the MCT predictions, was argued to be $d_{uc}=8$ \cite{MCT-spinodal,biroli-bouchaud}, as indeed also proposed for disordered spinodals \cite{dahmen-sethna}. The theoretical understanding of the spinodal of the RFIM in finite dimensions is therefore the crucial missing step to conclusively assess the physical content of the MCT and the status of the dynamical transition predicted by  the MF theory of glass-forming liquids.\\
It is well known however that rigorously speaking a spinodal cannot in general exist in finite dimensions, even above the upper critical dimension. This is due to the presence of thermal fluctuations that destroy metastability through nucleation phenomena. What could at best remain are therefore vestiges of the singularity in the form of a crossover behavior (as,  \textit{e.g.}, seen between spinodal decomposition and nucleation-and-growth in demixing transitions). However, in several cases, and for the RFIM, the fluctuations induced by the quenched disorder play a central role. The long-distance physics can then be directly studied at zero temperature \cite{nattermann}. This in particular implies that a spinodal can now be \textit{rigorously} defined beyond MF.\\
In the following, we focus on the RFIM and present a complete theory of its zero-temperature spinodal in all dimensions. 
The Hamiltonian of the model reads
\begin{equation}
\label{eq_hamiltonian_RFIM}
\mathcal H=-J\sum_{<ij>}S_iS_j-\sum_i[h_i+H(t)]S_i
\end{equation}
where Ising spins $S_i=\pm 1$ are placed on the vertices of a $d$-dimensional hypercubic lattice, $<ij>$ denotes distinct nearest-neighbor pairs, the $h_i$'s are random fields independently taken from a Gaussian distribution of zero mean and variance $\overline{h_i^2}=R^2$, and $H$ is the external magnetic field. The coupling $J$ is taken as the unit of energy and temperature ($J,k_B\equiv 1$).\\
Studying the zero-temperature spinodal of the RFIM means following its hysteretic, metastable behavior: one changes $H$ quasi-statically and let the system evolve under a $T=0$ Metropolis dynamics: a spin is flipped as soon as the local field at its site, $\sum_{j\backslash i}S_j+h_i+H$, changes sign, whatever the location of this spin \cite{sethna93,perkovic-sethna} [see also the supplemental material (SM) \cite{suppmat}]. 
There is a critical value of the disorder $R=R_c$ that separates a high-disorder regime where the hysteresis loop in the magnetization-versus-field plane is continuous and a low-disorder regime where a macroscopic magnetization jump is observed \cite{sethna93}. In the former, the avalanches, which are discontinuous changes of the system's configurations and are typical of the evolution of disordered systems at $T=0$, are of limited size whereas in the latter an infinite, system-spanning and compact, avalanche is found. For $R<R_c$, the value of the applied field (the ``coercive field'') at which the system spanning avalanche takes place can then be considered as the spinodal transition of the model: It is the limit of stability of the positively magnetized phase when the field is decreased or of the negatively magnetized one if the field is increased. (For $R>R_c$, there is no transition at all.)\\
The key questions we  address are: Under what conditions, if any, does this well-defined RFIM spinodal, which is in the same universality class as the MCT dynamical transition of glass-forming liquids, display critical behavior? What is the physical mechanism governing the spinodal transition and its long-distance universal behavior?\\
A naive response, supported by some numerical work \cite{dahmen-sethnaunpublished}, is that the spinodal should be MF-like and critical, \textit{i.e.}, with diverging correlation length and susceptibility, above an upper critical dimension $d_{uc}=8$ and noncritical, \textit{i.e.}, with finite correlation length and susceptibility, below. By combining analytical and numerical investigations, we show that this is not what happens. The critical behavior predicted by MF theory is absent {\it in any finite dimension}, even above $d_{uc}=8$. 
It is destroyed not only by thermal fluctuations, which was already well known, but also by rare athermal nonperturbative 
fluctuations. There is still a transition at $T=0$, but it greatly differs from its MF counterpart.

\begin{figure}[h]
\includegraphics[width=\linewidth]{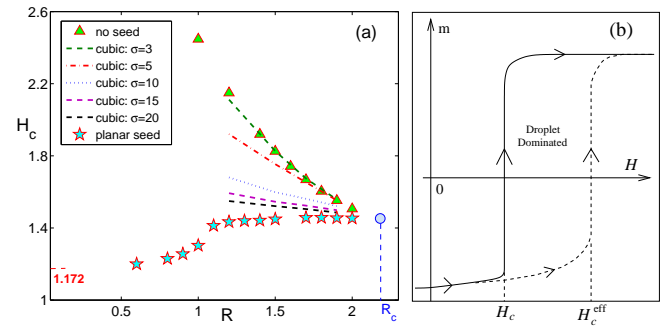}
\caption{(a) Coercive field $H_c$ versus disorder strength $R$ for the $3d$ driven RFIM at $T=0$ (simulation results). Triangles: full PBC. Stars: full PBC with a preexisting interface. Dotted and dashed lines: full PBC with a cubic seed of side $\sigma$. 
(b) Sketch of the ascending branch of the hysteresis loop in the magnetization-field plane; the exact (full line) and effective (dashed line) coercive fields of (a) are also indicated.}
\label{fig1a}
\end{figure}

We start with a surprising outcome of computer simulations for $3$-dimensional cubic lattices and periodic boundary conditions (PBC). For $R<R_c\simeq 2.2$ a spinodal takes place, but the value of the coercive field, $H_c(R)$, increases as one decreases the disorder below the critical value $R_c$: see for instance Fig. 1 of Ref.~[\onlinecite{perkovic-sethna}], Figs. 8 and 16 of  Ref.~[\onlinecite{perez-reche}], Fig. 1 of Ref.~[\onlinecite{liu-dahmen}], and the triangles in Fig. \ref{fig1a}a (obtained for linear size $L=90$ and averaging over $2400$ samples).  This is counterintuitive as one expects the opposite trend, with weaker disorder in general reducing the width of the hysteresis loop \cite{footnote-faceted}. This is even more puzzling if one studies the $R\rightarrow 0$ limit. In this case one can show (see below) that $H_c(0^+) \le 2(2-\sqrt 2)\approx 1.172$, which is less than $H_c(R_c)\approx 1.45$. Therefore the numerical curves appear to head toward the wrong limit for $R\rightarrow 0$ (see Fig. \ref{fig1a}a).\\
The solution of this quandary is in the role played by rare events which induce extremely strong finite-size effects. The true coercive field $H_c(R)$ is actually {\it monotonically increasing} with $R$, contrary to what found in the above simulations. For $R$ strictly less than $R_c$ \cite{footnote_Rc} it coincides with the depinning threshold \cite{depinning} of an infinite surface along the most difficult depinning direction (on depinning, see Refs. [\onlinecite{nattermann,reviewdepinning1,reviewdepinning2}]). To see this, consider a field $H$ just slightly larger than the threshold value for the depinning transition (henceforth we focus on the ascending branch of the hysteresis loop: see Fig. 1b). In this regime compact droplets of up spins of linear size $r$ larger than the longitudinal correlation length for a pinned interface, $\xi_{\parallel}(H)\sim (H-H_c^{dep})^{-\nu_{dep}}$, are unstable. Their interface ballistically expands, resulting in the \textit{unbounded} growth of the associated droplets. This can be shown by estimating the probability $p_{\exp}( r)$ that a droplet can expand to infinity. By decomposing it into independent contributions, each one of them related to the probability of expansion over a length much larger than the transverse correlation length, $\xi_{\perp} \sim \xi_{\parallel}^\zeta$ with $\zeta$ the roughness exponent, one finds: $p_{exp}( r)=\prod_{n=0}^{\infty} p(r_n)$ where $r_n=r+ n K \xi_{\perp}$ with $K \gg1$. This product converges to a finite value as long as $[1-p( x)] \to 0$  faster than $1/x$ when $x \to \infty$, which is certainly the case (by using large-deviation analysis, valid for $x\gg1$, we actually predict an exponential decrease). Since the medium of down spins is not characterized by any long-range correlation besides its magnetic order, we expect that these unstable droplets form a dilute gas  with a density $\rho\sim e^{-c r^d}$ where $c$ is a constant of order 1. The smallest ones among them, but still unstable with probability $O(1)$, have a linear size of the order of  $\xi_\parallel$ and hence are at a typical distance $\ell(H)\sim e^{c\xi_\parallel^d/d}$. Due to their ballistic expansion, they invade the system and induce a transition to the positively magnetized (stable) state in a time
\begin{equation}\nonumber
\tau(H)=\frac{\ell(H)}{v(H)}\propto(H-H_c^{dep})^{-\beta} \exp \left(c' [H-H_c^{dep}]^{-d\nu_{dep}}\right),
\end{equation} 
where $v\propto(H-H_c^{dep})^\beta$ is the velocity of an unpinned interface close to the transition \cite{nattermann} and $c'$ is a constant of order one. This implies that as long as $H>H_c^{dep}$ the negatively magnetized state is unstable. Since we do not expect more efficient destabilizing processes in this disorder regime, we obtain that $H_c=H_c^{dep}$, \textit{i.e.}, the coercive field coincides with the depinning threshold of an infinite interface. This conclusion is similar to that reached in Ref. [\onlinecite{muller}] from field-theoretical arguments.\\
We can complete this theoretical analysis by addressing the issue of the potential criticality  for $H\rightarrow H_c( R)^-$. For concreteness, we focus on the susceptibility $\chi=\partial m/\partial H$ and analyze whether it diverges when approaching the spinodal. When changing the applied field from $H$ to $H+\delta H$ below the depinning threshold, the only critical behavior is due to droplets of size $r \gtrsim \xi_{\parallel}(H)$ that expand through avalanches over a scale $\xi_{\perp}(H)$. From the known behavior near a depinning transition \cite{reviewdepinning1,reviewdepinning2}, each portion of linear size $\xi_{\parallel}(H)$ of the boundary of a droplet should advance over a volume that in average scales as $(\xi_{\parallel}^{d-1}\xi_{\perp})^{2-\tau}$, where $\tau$ is the critical exponent of the avalanche-size distribution.  As $\xi_{\parallel}$ and $\xi_\perp$ diverge when $H \rightarrow H_c^{-}$, avalanches take place on all scales. However their density is exponentially suppressed since the droplet probability goes as $\exp(-c r^d)$ (see above). 
This leads to a singular contribution akin to that found in Griffiths phases \cite{griffiths}: Thermodynamic quantities are singular but display only essential singularities and do not diverge at the spinodal. For instance, the susceptibility should vary as
\begin{equation}
\begin{aligned}
\label{eq_susceptibility}
\chi(H) - \chi_{reg}(H) &\propto \int_{r \gtrsim \xi_{\parallel}(H)} dr \, (\xi_\parallel^{d-1}\xi_{\perp})^{2-\tau}  \left(\frac{r}{\xi_{\parallel}}\right)^{d-1}  \,e^{-c r^d}  \; \\&
\propto  (H_c-H)^{-\omega} e^{-c' (H_c-H)^{-d\nu_{dep}}}
\end{aligned}
\end{equation}
where $\omega=\nu_{dep}[(d-1+ \zeta_{dep})(2-\tau)+1]$, $\chi_{reg}$ denotes a regular, finite, contribution; $\chi(H)$ thus goes to a finite value when $H \rightarrow H_c^{-}$, albeit in a nonanalytical way.

We are now in a position to explain the simulation results and the role played by finite-size effects. For system sizes $L\gg \ell(H)\sim e^{c\xi_\parallel(H)^d/d}$, a droplet of linear size at least $\xi_{\parallel}$ is typically present in the system at the applied field $H$ and the physics is controlled by its (ballistic) expansion. On the other hand, for $L\ll \ell(H)$, no unstable droplets are present and the negatively magnetized state is stable even though $H>H_c$. Depinning is in some sense postponed as it requires larger applied fields. Due to the extremely fast rise of $\ell(H)$, as soon as $\xi_\parallel$ starts to grow, one observes in practice a ``transition'' at a field $H_c^{eff}(L)$, which is defined by $\ell(H_c^{eff}(L)) \sim L$. Its $L$-dependence is only logarithmic, $H_c^{eff}(L)\simeq H_c+ \tilde c \, (\ln L)^{-\frac 1{d\nu_{dep}}}$, and is therefore hard to detect.\\
To confirm the above interpretation, we have first studied  the RFIM at $T=0$ on a cubic lattice of size $L$ by computer simulation (with $20\leq L\leq 60$ and full PBC; details in the SM \cite{suppmat}), but now starting from an initial condition with a plane of up spins when all other spins in the bulk are down. This amounts to studying the depinning of an infinite interface \cite{depinning}. The resulting coercive field $H_c( R)$ corresponds to the bottom curve in Fig. \ref{fig1a}a. The presence of a preexisting interface lowers the value of the coercive field. The latter is now found to decrease with decreasing disorder strength. (Note that, except in the vicinity of the critical endpoint, $R_c\approx 2.2$, where bulk and interface mechanisms also become more and more intertwined, the results are weakly dependent on system size for $L\geq 30$.) On the basis of our previous arguments, the $H_c$ line that we now find is thus the correct spinodal transition line.\\
To further investigate the role played by droplets in inducing finite-size effects, we have run simulations of the model with full PBC in the presence of an initial seed (a droplet) in the form of a cube of up spins in the bulk of down spins. As with the preexisting interface, by doing this we put by hand the rare events needed to induce the transition. We have then determined the effective coercive field $H_c^{eff}(\sigma,R)$ at which a macroscopic avalanche first takes place as a function of the cube side $\sigma$ and of the disorder $R$ (for  $L=60$, see SM \cite{suppmat}). The results are shown in Fig. \ref{fig1a}. The curves interpolate between the full PBC in the absence of seed and the transition line with a preexisting interface. They can be considered as ``iso-cube-size'' crossover lines: They would be observed in samples of linear size $L$ containing droplets no larger than $\sigma$. On the basis of our above arguments, the sharp crossover that resembles a transition then takes place when $\xi_{\parallel}\sim \sigma$, which leads to: $H_c^{eff}(\sigma) \simeq H_c+ b \sigma^{-1/\nu_{dep}}$ [for $\sigma\ll L\ll \ell(H_c^{eff})$]. We have checked that this law is indeed compatible with our numerical data: see the SM \cite{suppmat}. Our numerical results further indicate that the typical values of $L$ reachable in simulations ($L\sim 100$) correspond to droplet sizes of a few units $\sigma \lesssim 3$, as we also find by direct inspection.

\begin{figure}[t]
\begin{center}
\includegraphics[width=0.9\linewidth]{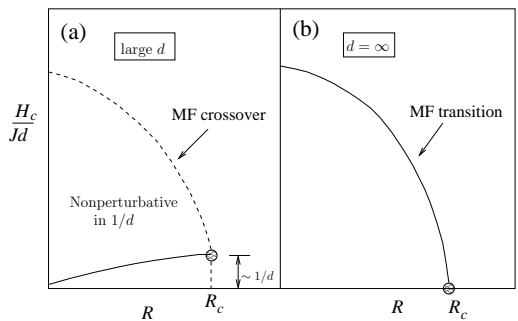}
\caption{Coercive field $H_c/(Jd)$ versus disorder strength $R$ for the driven RFIM at $T=0$ in a large but finite $d$ (a) and for $d= \infty$ (b). (a): Long dashed: Crossover associated with small droplets having $\sigma \sim 1$. Full: Actual transition associated with depinning and rare large droplets. (b) Full: Actual transition induced by purely local droplets. Compare also with Fig. 1a.}
\label{fig3}
\end{center}
\end{figure}

Having clarified the critical behavior and the finite-size effects in $d=3$, we turn to the issue of the potential criticality in high dimension and the relation with MF results. On the fully connected lattice and on Bethe lattices \cite{Connect3} the susceptibility $\chi=\partial m/\partial H$ diverges when approaching the spinodal as $[H_c( R)-H]^{-1}$. This is associated with a scale-free, power-law, distribution of avalanches (with the same exponent $\tau=3/2$ as at the critical endpoint $R_c$ \cite{dahmen-sethna}). Perturbative arguments suggest that this behavior persists as long as $d\ge d_{uc}=8$, but our results which hold in any $d$ invalidate this conclusion. \\
What is the effect of the rare events discussed above in high dimension?
Considering the $R\rightarrow 0^+$ limit is particularly instructive. The instability due to the depinning and the subsequent expansion of rare droplets implies that the coercive field in a hyper-cubic lattice in $d$ dimensions is less than or equal to the threshold for the depinning of a $(d-1)$-manifold (along the most difficult direction) in $d$ dimensions: $H_c^{(d,RFIM)}( 0^+)\le H_c^{(d,dep)}( 0^+)$. Moreover, a sufficient condition for depinning is that an infinite hyper-plane of $+1$ spins in a bulk of $-1$ spins is able to move by one lattice spacing. Consider now the adjacent hyper-plane that the infinite manifold has to invade. The spins belonging to it exactly form a $(d-1)$-dimensional RFIM. If the applied field is larger than the coercive value of this model, $H_c^{(d-1,RFIM)}(0^+)$, then the configuration flips to the positively magnetized state and the manifold indeed moves. This implies $H_c^{(d,dep)}(0^+)\le H_c^{(d-1,RFIM)}(0^+)$ and, in consequence, $H_c^{(d,RFIM)}(0^+)\le H_c^{(d-1,RFIM)}(0^+)$. The conclusion is that whatever $d$, $H_c^{(d,RFIM)}(0^+)\le H_c^{(d=2,dep)}( 0^+)=2(2-\sqrt 2)\approx 1.172$ \cite{2d-limit}. On the other hand, the analysis of the RFIM when $d= \infty$ (fully connected lattice) \cite{dahmen-sethna} leads to a coercive field that increases as $R$ decreases and goes to a value $\propto d \to \infty$ when $R\to 0^+$! The outcome is sketched both for a large $d$ and for $d= \infty$ in Fig. \ref{fig3} after having rescaled the coupling $J$ by $1/d$. \\
Clearly MF theory and the limit of infinite dimensions get the critical behavior and even the transition line completely wrong. The droplet-induced avalanches are nonperturbative in $1/d$ and are not taken into account within MF treatments, which instead capture only avalanches triggered by a single spin flip. (Think for example of a Bethe lattice where flipping spins on one branch does not influence the propagation of an avalanche on another branch.) From the perspective discussed above in terms of cubic seeds, the MF description corresponds to $\sigma=1$: the predicted critical line therefore corresponds to a crossover, the sharper the larger $d$ and whose signature can be seen in simulations with limited system sizes as for $d=9$ in Ref. [\onlinecite{dahmen-sethnaunpublished}], but not to a transition. The actual transition is located in a very different region of the parameter space. In the intermediate droplet-dominated regime (see Fig. \ref{fig1a}a) both the time-scale to evolve from one magnetic state to the other and the distance between droplets diverge when $d\rightarrow \infty$: This regime is non-perturbative in $1/d$. (Note that there is no sharp change of the  crossover in $d=8$, but that the perturbative fluctuations may nonetheless have a stronger influence below this dimension.)

What are the consequences of our findings on the current understanding of glassy dynamics? On the basis of the mapping that relates the dynamical transition of glassy systems to the spinodal of the RFIM \cite{MCT-spinodal}, we can conclude that the physics described by the MCT is one of avalanches of motion created by a localized rearrangement. (The analog for glassy dynamics of a spin flip in the RFIM is a localized relaxation event.) From this point of view there is a clear connection with the phenomenon of dynamical facilitation \cite{KCM,CG,simulations}. With this in mind, it would be worth directly establishing from simulations and experiments whether facilitation and avalanches of motion are also generated, before thermally activated processes take over, by rare droplets whose size grow as one lowers the temperature.    
Finally we stress that our results imply that even when activated events are neglected the  MCT singularity can only have the meaning of a crossover, which is sharper the higher $d$. This has also been recently argued by Rizzo \cite{rizzo}, but contrary to his conclusion we find that disorder does not destroy the spinodal transition altogether; yet, it drastically changes its nature and its location via nonperturbative phenomena.  The MCT description then becomes invalid whenever thermal  activation {\it or} droplets with $\sigma \gg1$ take over \cite{footnoteappear}.\\
We conclude by noting that the instability or failure of a metastable state in the presence of quenched disorder is a quite general phenomenon with applications from materials to social science. The implication of our results 
on the role of rare relaxation events for these cases is certainly an issue worth being addressed in the future. 

\begin{acknowledgments}
We thank K. Dahmen, P. Le Doussal, T. Nattermann, G. Parisi, A. Rosso, J. Sethna and C. Toninelli for fruitful discussions. We acknowledge support from the ERC grant NPRGGLASS. Part of this work was  done during the 2014 KITP workshop on Avalanches, Intermittency, and Nonlinear Response in Far-From-Equilibrium Solids.
\end{acknowledgments}

\onecolumngrid

\newpage
\appendix

\section*{Supplemental Material: Spinodals with Disorder: from Avalanches in Random Magnets to Glassy Dynamics}

\twocolumngrid

In this Supplemental Material we first discuss the effect of the chosen zero-temperature dynamics, then give some details on the simulation, and finally expand on the results that we have obtained with the full boundary conditions and by introducing cubic seeds in the initial state of the system.

\subsection{Zero-temperature dynamics}
The quasi-statically driven RFIM at zero temperature has been studied with two different dynamics. 
One was introduced by Sethna and coworkers \cite{sethna93_sm,perkovic-sethna_sm} and corresponds to the 
zero temperature limit of the Metropolis dynamics. Robbins and coworkers \cite{ji-robbins_sm,koiler-robbins_sm} introduced another one: a domain-wall motion corresponding to a domain-growth dynamics, in which the spins may flip under the same condition as discussed in the main text, but with the restriction that their location is at the interface between the domains of up spins and down spins; in this case, a preexisting interface is introduced at the initial time.

\begin{figure}[h!]
\begin{center}
\includegraphics[width=\linewidth]{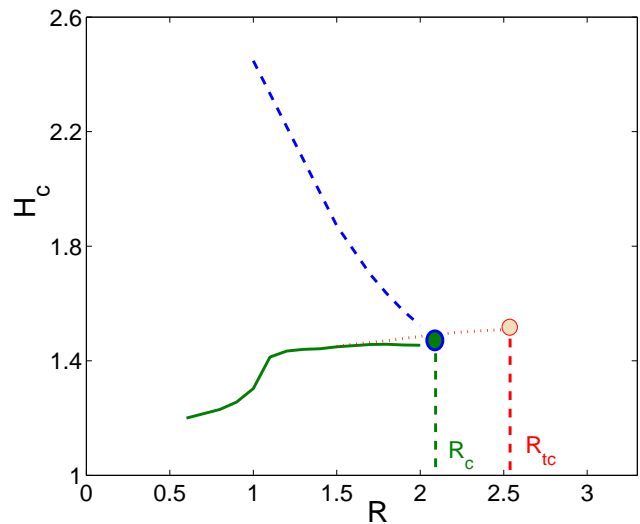}
\caption{(a) Coercive field $H_c$ versus disorder strength $R$ for the $3$-$d$ driven RFIM at $T=0$ (simulation results). In addition to our  data shown in Fig. 1a of the main text we have plotted as  a dotted line the depinning transition line obtained with the domain-growth dynamics in Ref. [\onlinecite{koiler-robbins_sm}]. Long dashed line: our data with full PBC (L=90). Full line: our data with full PBC and a preexisting interface.}
\label{fig4_sm}
\end{center}
\end{figure}

In the domain-growth dynamics, the critical field is that of the depinning of the interface. For the $3-d$ RFIM with gaussian random fields studied in the present work, Koiler and Robbins \cite{koiler-robbins_sm} found that below a (tri)critical value $R_{tc}\approx 2.52$, the growing domain is compact (the magnetization $m$ therefore jumps discontinuously at the transition) and the interface is self-affine whereas above $R_{tc}$ one finds a self-similar percolation-like mechanism that results in a continuous evolution of $m(H)$. The results are displayed together with those of the present study in Fig. \ref{fig4_sm}.  Note that the critical/coercive fields for the two dynamics seem to coincide within numerical accuracy below a disorder strength $R\approx 1.5$. Above this value, the Sethna \textit{et al.} dynamics allows nucleation events in the bulk and the preexisting interface therefore moves in a more favorable environment, with a reduced effective disorder. In consequence, depinning takes place at a lower value of $H$ than in the domain-growth dynamics. As one approaches the critical endpoint $R_c\approx 2.2$, bulk and interface mechanisms become more and more intertwined. As already pointed out \cite{sethna93_sm,dahmen-sethna_sm,perkovic-sethna_sm}, the critical behavior near $R_c$ is then different from that of the domain-growth dynamics around $R_{tc}$. 

\subsection{Simulation details}

In our simulation, to accelerate the dynamics we have used a faster-than-the-clock algorithm \cite{krauth_sm}. We first look at the spins that are unstable. We use a rejection-free algorithm in which we sample the random time interval for the next spin to be flipped. The waiting time for flipping an unstable spin is a Poisson variable whose mean is determined by the fraction of unstable spins. We then choose a spin at random from the unstable ones and flip it with probability one. We next re-compute the local fields on the neighboring spins and update the list of unstable spins. Since this is a local process, the computation cost is small and the algorithm is much faster than with the standard Metropolis one. We have checked that the output obtained in this way is the same as that obtained with the plain Metropolis algorithm.

We have investigated three types of geometry and initial conditions for the $3$-$d$ RFIM on a $L\times L\times L$ cubic lattice with periodic boundary conditions (PBC) in all directions: (1) starting with all spins down, (2) starting with a preexisting plane of up spins at half-height in the sample and a bulk of down spins, and (3) starting with a cubic seed of up spins in a bulk of down spins.

For case (1), we have considered random-field strength $R$ from $1$ to around $2.2$, the latter corresponding to the critical point which we have checked from finite-size scaling analysis. We have taken $L$ from $20$ to $120$ and a number of samples $\mathcal N$  ({\i.e.}, realizations of the random fields) depending on the value of $R$: For $R=1$, $\mathcal N=2400$  for $L=80,90,100$ and $\mathcal N=6000$ for smaller values of $L$. For $1<R\le 1.5$ we used $L=90$ and $\mathcal N=2400$.
For larger values of $R$ a smaller number of samples, typically $1200$ was found sufficient. Note that even with the faster-than-the-clock algorithm, a simulation for $R=1$ and $L=120$ takes of the order of two days on a 2.6 GHz computer.

For the case (2) we have taken $L$ from $20$ to $60$ with $\mathcal N=2400$ for the smaller systems and $1200$ for the larger ones.

Finally, for case (3), we first checked that finite-size effects due to a too small ratio between the system size and the seed size are negligible when $L/\sigma\le 3$. We have then studied $\sigma$ from $3$ to $20$ with $L=60$ and $\mathcal N=1200$.

Note that the intrinsic simulation errors are very small and always less than the symbols used in the figures
\subsection{Data Analysis}
\begin{figure}[h!]
\begin{center}
\includegraphics[width=\linewidth]{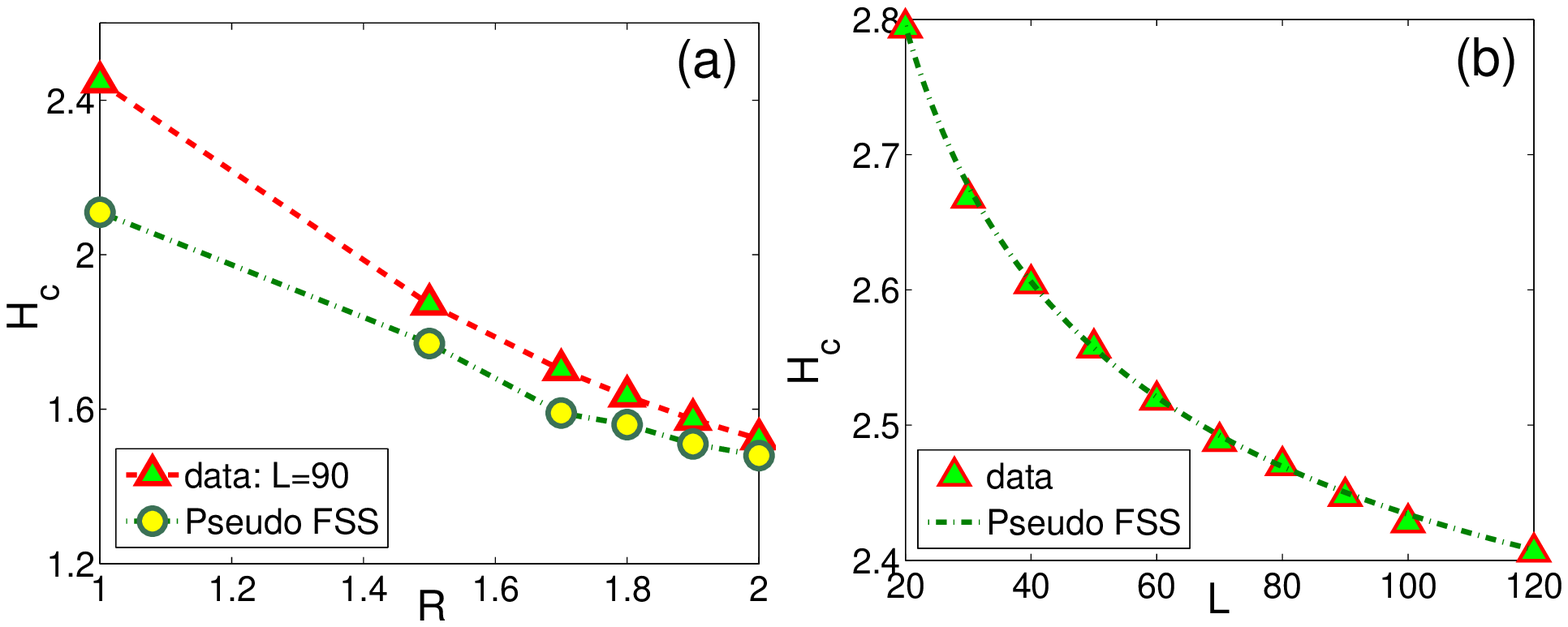}
\caption{Effective coercive field $H_c$ for the $3$-$d$ driven RFIM at $T=0$ with full PBC and no preexisting plane or seed. (a) $H_c$  versus disorder strength $R$. The triangles correspond to $L=90$ and the circles to the result of a pseudo finite-size scaling analysis illustrated in (b). (b) Fit of $H_{c,L}$ for $R=1$ and $L$ from $20$ to $120$ to the form $H_{c,L}=H_c^{eff}+A L^{-x}$ with $H_c=2.11$, $A=2.78$, and $x=0.47$. }
\label{fig2_sm}
\end{center}
\end{figure}

When the system is studied with full boundary conditions [case (1)], even with system sizes as large as $L=120$ one observes an effective coercive field as exponentially rare droplets cannot be found in practice. In Fig. \ref{fig2_sm}a we show $H_c( R)$ obtained for $L=90$ as well as the result of a pseudo finite-size scaling analysis in which we fit the effective coercive field for each value of the random-field strength $R$ to the form $H_{c,L}=H_c^{eff}+AL^{-x}$. As seen from Fig. \ref{fig2_sm}b the fit is good over the range of $L$ studied (from $20$ to $120$) and gives values of $H_c^{eff}$ comparable to those obtained for the largest values of $L$. The pseudo finite-size scaling procedure should however break down for (much) larger values of $L$ since $H_c^{eff}$ is not actually constant but rather decreases  logarithmically with $L$, as discussed in the main text.

\begin{figure}[h!]
\begin{center}
\includegraphics[angle=-90,width=\linewidth]{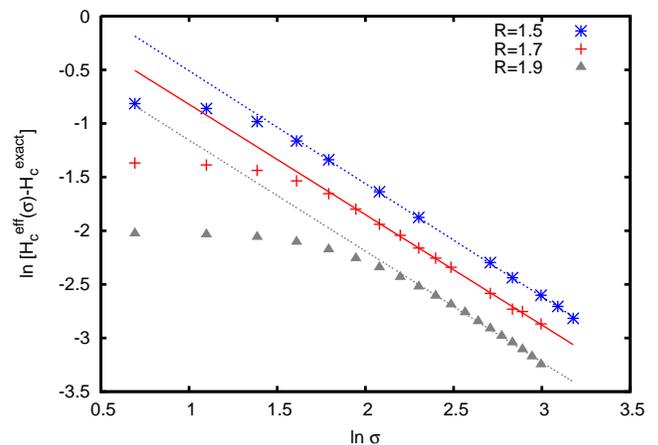}
\caption{$T=0$ driven RFIM in $d=3$ with a cubic seed of side $\sigma$. The system size is $L=60$. Difference with the coercive field in the presence of a preexisting interface as a function of $\sigma$ on a log-log plot. The slope of the curves is about 1.05, within $20 \%$ of the estimated value of $1/\nu_{dep}\approx 1.14-1.3$ \cite{ledoussal-chauve_sm}.}
\label{fig3_sm}
\end{center}
\end{figure}

As explained in the main text, in the case (3) where a cubic seed is present, we have studied the variation of the effective coercive field $H_c^{eff}(\sigma,L, R)$ with the cube side $\sigma$ for a large system size $L=60$ and several values of the disorder strength $R$. The difference $H_c^{eff}(\sigma,L, R)-H_c^{(\infty)}( R)$ is plotted versus $\sigma$ on a log-log scale in Fig. \ref{fig3_sm}. For  $\ln \sigma \lesssim 1.6$ the effective coercive field is essentially constant and is given by the value in the absence of seed and for $\sigma\sim L$  one has severe finite-size effects. However, one observes in between a power law. The effective exponent is smaller but roughly compatible, considering the uncertainty of the present determination, with $1/\nu_{dep}$ in $d=3$: we find $1/\nu_{dep}\approx 1.05$ whereas the expected value is $1/\nu_{dep}\approx 1.14-1.3$.\cite{ledoussal-chauve_sm} 

\begin{thebibliography}{99}
\bibitem{footnote-faceted}
This applies at least when there is no faceted growth of an interface: see Ref. [\onlinecite{ji-robbins}].
\bibitem{depinning}
We define the depinning threshold with respect to a dynamics that also allows spins in the bulk to flip. 
This is different from the one used in Refs. [\onlinecite{ji-robbins},\onlinecite{koiler-robbins}], where only spins close to the interface are allowed to flip. In the regime of self-affine growth both dynamics are expected to be in the same universality class 
but to display different critical/coercive fields, see the SM \cite{suppmat}.
\bibitem{Connect3}
In order to display a spinodal the connectivity must be strictly larger than 3 \cite{dhar97}.
\end{thebibliography}
\end{document}